\documentclass[prl,twocolumn,preprintnumbers,amsmath,amssymb,superscriptaddress]{revtex4-1}

\usepackage{graphicx}
\usepackage{bm}
\usepackage{color}
\usepackage{textcase, natbib, url, ulem}
\usepackage[colorlinks=true, linkcolor=blue, urlcolor=blue, citecolor=blue]{hyperref}
\usepackage{amsmath,amssymb}
\usepackage{mathtools}

\definecolor{darkblue}{rgb}{0,0,0.5}
\definecolor{darkgreen}{rgb}{0,0.5,0}
\definecolor{darkred}{rgb}{.7,0,0}
\definecolor{purple}{rgb}{0.5,0,0.6}
\definecolor{orange}{rgb}{1,0.5,0}
\definecolor{grey}{rgb}{.6,.6,.6}
\definecolor{lightpink}{rgb}{1,0.7,0.75}
\definecolor{pink}{rgb}{1,0.4,0.58}
\definecolor{deeppink}{rgb}{1,0.08,0.58}

\begin{document}


\title{Mesoscopic phase behavior in a quantum dot  around crossover between single-level and multi-level transport regimes}


\author{S. Takada}
\affiliation{Department of Applied Physics, University of Tokyo, Bunkyo-ku, Tokyo, 113-8656, Japan}
\affiliation{Univ. Grenoble Alpes, Inst NEEL, F-38042 Grenoble, France}
\affiliation{CNRS, Inst NEEL, F-38042 Grenoble, France}
\author{M. Yamamoto}
\affiliation{Department of Applied Physics, University of Tokyo, Bunkyo-ku, Tokyo, 113-8656, Japan}
\affiliation{PRESTO, JST, Kawaguchi-shi, Saitama 331-0012, Japan}
\author{C. B\"{a}uerle}
\affiliation{Univ. Grenoble Alpes, Inst NEEL, F-38042 Grenoble, France}
\affiliation{CNRS, Inst NEEL, F-38042 Grenoble, France}
\author{A. Ludwig}
\affiliation{Lehrstuhl f\"{u}r Angewandte Festk\"{o}rperphysik, Ruhr-Universit\"{a}t Bochum, Universit\"{a}tsstra\ss e 150, 44780 Bochum, Germany}
\author{A. D. Wieck}
\affiliation{Lehrstuhl f\"{u}r Angewandte Festk\"{o}rperphysik, Ruhr-Universit\"{a}t Bochum, Universit\"{a}tsstra\ss e 150, 44780 Bochum, Germany}
\author{S. Tarucha}
\affiliation{Department of Applied Physics, University of Tokyo, Bunkyo-ku, Tokyo, 113-8656, Japan}
\affiliation{RIKEN Center for Emergent  Matter Science (CEMS), 2-1 Hirosawa, Wako-shi, Saitama 31-0198, Japan}


\date{\today}

\begin{abstract}
The transmission phase across a quantum dot (QD) is expected to show a \textit{mesoscopic} behavior, where the appearance of a phase lapse between Coulomb peaks (CPs) as a function of the gate voltage depends on the orbital parity relation between the corresponding CPs.
On the other hand such a mesoscopic behavior has been observed only in a limited QD configuration (few electron and single-level transport regime) and \textit{universal} phase lapses by $\pi$ between consecutive CPs have been reported for all the other configurations.
Here we report on the measurement of the transmission phase across a QD around the crossover between single-level and multi-level transport regimes employing an original two-path quantum interferometer.
We find a mesoscopic behavior for the studied QD.
Our results show that the universal phase lapse, a long standing puzzle of the phase shift, is absent for a standard QD, where several tens of successive well-separated CPs are observed.
\end{abstract}

\pacs{72.10.Fk, 72.20.Dp, 73.23.Hk, 85.35.Ds}


\maketitle


Quantum coherent transport is essentially different from its classical counter part.
The quantum phase of an electron produces various kinds of quantum interference phenomena, such as the Aharonov-Bohm (AB) effect, weak localization and universal conductance fluctuations.
Characterization of the phase is therefore required to fully describe the coherent transport.
One of the most fundamental problems is the scattering phase through an (artificial) atom or a quantum dot containing electrons.
That can be studied by employing a quantum two-path interferometer.
Indeed transmission phase shift of an electron across a quantum dot (QD) was measured by embedding a QD into one arm of a multi-terminal AB interferometer \cite{Schuster1997}.
It was confirmed that the phase evolves by $\pi$ across a Coulomb peak (CP), where the number of electrons inside the QD changes by $1$, as expected from Friedel's sum rule \cite{Yeyati1995}.
On the other hand, unexpected abrupt $\pi$ phase lapses were found between all successive CPs irrespective of various parameters of the QD.
Due to its robustness such a phase behavior is termed \textit{universal}.
Later, the transmission phase shift across a few electron QD was also investigated \cite{Heiblum_phase_2005}.
In this experiment for a QD containing up to 10 electrons the phase showed phase lapses between some CPs while a smooth phase shift between other CPs was observed.
Such a phase behavior, where the appearance of the phase lapse depends on parameters of the mesoscopic system, is expected theoretically \cite{Lee1999} and is termed \textit{mesoscopic}.
However for the larger QD configuration containing more than $14$ electrons the universal phase behavior was recovered.
This observation invokes the potential importance of the crossover between single-level ($\Gamma < \delta$) and multi-level ($\Gamma > \delta$) transport regimes, where $\Gamma$ is the level broadening and $\delta$ is the single-level spacing of a QD. Larger quantum dots generally have smaller level-spacing, which might lead a crossover between the two regimes within the studied energy scale.

It has been shown theoretically that the presence of a phase lapse is related with the symmetry of orbital wave functions.
A phase lapse is expected to appear only in the valley between CPs with the same parity of orbital wave functions, leading to the mesoscopic behavior \cite{Lee1999}.
On the other hand, a generic explanation of the universal regime is still under debate despite many theoretical works devoted to it \cite{Baltin1999, Silvestrov2000, Yeyati2000, Hackenbroich2001, Kim2002a, Golosov2006, Bertoni2007a, Karrasch2007, Goldstein2009a, Molina2012}.
Theoretical results failed to reproduce such a universal regime observed previously \cite{Schuster1997, Heiblum_phase_2005}.
One difficulty for the understanding of the phase shift lies in the fact that only a few experimental works exist \cite{Schuster1997, Heiblum_phase_2005, Sigrist2004, Aikawa2004B}.
Indeed it is not easy at all to measure the \textit{true} transmission phase shift of an electron in mesoscopic systems due to boundary conditions imposed by the contacts  \cite{Yeyati1995, Yacoby1996}.
The measured phase shift can be unintentionally modified from the \textit{true} phase shift by contributions from multi-path interferences \cite{Entin2002, Simon2005}.
Recently we have demonstrated a way how to measure the \textit{true} transmission phase of an electron \cite{TakadaAPL2015}.
We have shown that the criterion used in our experiments using our original two-path interferometer ensures the proper measurement \cite{TakadaAPL2015} while those used in other previous experiments are a bit less reliable \cite{Schuster1997,Heiblum_phase_2005}.
Therefore a careful experimental investigation of is required to have comprehensive understanding of the phase behavior across a QD.

Here we investigate the transmission phase across a QD around the crossover between single-level and multi-level transport regimes using our original interferometer \cite{TakadaAPL2015, Yamamoto:2012fk, Tobias2014, Aharony2014}.
In contrast to the previous experiments \cite{Schuster1997, Heiblum_phase_2005} we observe mesoscopic phase behavior with a QD that is not in the few electron regime both in the single-level and multi-level transport regimes.
When the QD is made larger, an overlap between adjacent CPs starts becoming larger, which prevents us from observing a clear phase shift of $\pi$ as well as the occurrence of the phase lapse.
Eventually we do not observe the universal phase behavior for the maximum size of the QD accessible.
Finally we show the asymmetric phase behavior observed in the high temperature Kondo regime ($T \gg T_{\rm K})$, which supports that the mesoscopic phase behavior is indeed related with the orbital parity relation between adjacent CPs.
The long standing problem of the universal phase behavior is absent in our QD.

\begin{figure}[htbp]
	\includegraphics[width=0.49\textwidth]{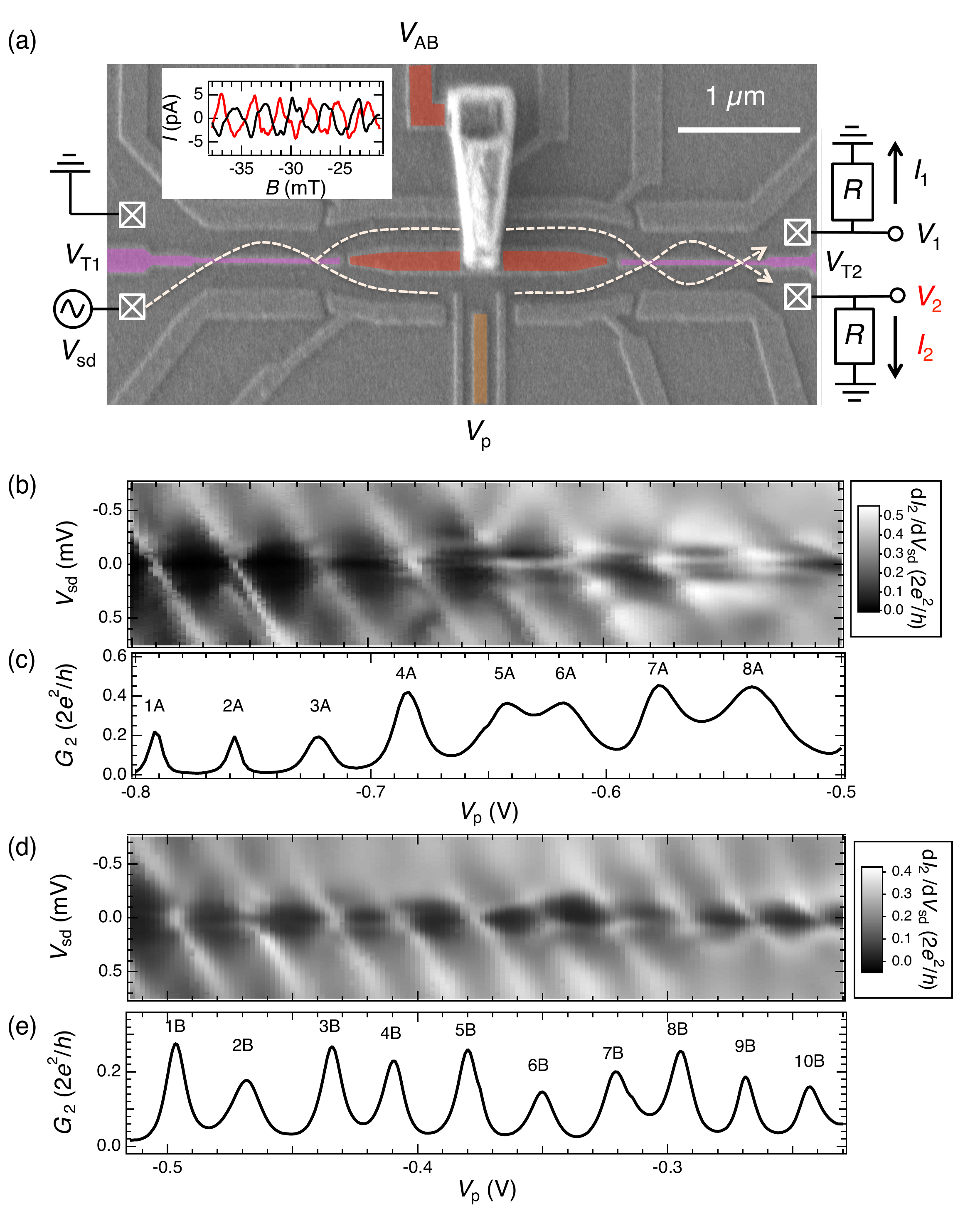}
	\caption{\label{fig:device}(a) A SEM image of the measured device and experimental setup. $R=10\ {\rm k\Omega}$ is used to convert the currents to voltages. The dashed lines indicate the typical trajectory of two paths. (Inset) Typical magneto oscillations of output currents $I_{\rm 1}$ (black) and $I_{\rm 2}$ (red). The smoothed background is subtracted and only the oscillating components are plotted. (b), (d) Coulomb diamonds for the Coulomb peaks in regime A and B. 	The number of electrons inside the quantum dot is decreasing towards more negative $V_{\rm p}$. (c), (e) Coulomb peaks measured at 0 dc bias voltage for the two regimes. The peaks are named as peak 1A to 8A for regime A and as 1B to 10 B for regime B.}
\end{figure}
Our quantum interferometer was fabricated from a GaAs/AlGaAs heterostructure that hosts a 2DEG with an electron density of $n=3.21 \times 10^{11}\, {\rm cm^{-2}}$, electron mobility of $\mu=8.6 \times 10^5\, {\rm cm^2/Vs}$ located 100 nm below the surface with modulation doping and a 45 nm spacer between doping and 2DEG.
It is defined by applying negative voltages on the metallic Schottky gates deposited on top of the substrate and depleting the 2DEG underneath [Fig.\,\ref{fig:device}a].
The interferometer consists of an Aharonov-Bohm (AB) ring at the center and tunnel-coupled wires at the both ends.
The AB ring is formed by the gate voltage $V_{\rm AB}$ applied through the metallic air bridge.
A QD is formed at the lower arm of the AB ring and its energy level is controlled by changing the plunger gate voltage $V_{\rm p}$.
Electrons are injected from the lower left contact by applying an AC bias [$10 \sim 20\ {\rm \mu V}$, 23.3 Hz] and currents are measured at the two contacts on the right side through voltage measurements across the resistances.
When the tunnel-coupled wires are tuned to half beam splitters, the interferometer works as a \textit{pure} two-path interferometer and shows anti-phase oscillations of the two output currents $I_{\rm 1}$ and $I_{\rm 2}$ [Fig.\,\ref{fig:device}a inset] \cite{Yamamoto:2012fk, Tobias2014, Aharony2014}

Firstly we deplete the electrons underneath both tunnel-coupled wires by applying large negative voltages on $V_{\rm T1}$ and $V_{\rm T2}$.
As a result the upper and lower parts of the device are electrically isolated.
All the injected current into the lower left contact passes through the QD and is recovered at the lower right contact.
This allows us to observe well-defined CPs and to characterize the QD.
We measure the phase in two different QD configurations referred to regime A and B.
Between the two regimes we changed $V_{\rm AB}$ by 50 mV, which significantly changes the shape of the QD and allows us to access a different size of the QD.
Fig.\,\ref{fig:device}b (d) shows the Coulomb diamonds in regime A (B), where the corresponding differential conductance is plotted in the plane of $V_{\rm p}$ and $V_{\rm sd}$.
The conductance as a function of $V_{\rm p}$ at $V_{\rm sd}=0$ in the regime A (B) is shown in Fig.\,\ref{fig:device}c (e).
We estimate the characteristic energy scales of the QD from the Coulomb diamonds.
The charging energy $U$ is gradually changed from $0.9 - 0.8$ meV for the smaller QD (regime A), while it varies from $0.9 - 0.5$ meV for the large QD (regime B).
Regime B represents a slightly larger QD configuration with a corresponding smaller charging energy.
However we tuned the gate voltages in such a way that the QD size at the right part of Fig 1b (regime A) is equivalent to the left part of Fig. 1d (regime B).
The single-level spacing $\delta$ is measured around the peak 2A and 1B for the regime A and B, respectively, and is about $0.2$ meV for both peaks, and decreases for more positive $V_{\rm p}$.
Given the parameters above, our QD is not in the few electron regime and the QD is expected to contain a few tens of electrons.

Another important feature found for the QD is the Kondo correlation \cite{Kondo1964,Gordon1998}.
In Fig.\,\ref{fig:device}b the differential conductance is enhanced around the zero source-drain bias for the valley between the peak 5A and 6A as well as the peak 7A and 8A.
This enhancement corresponds to the zero-bias anomaly and is a typical signature of the Kondo correlation.
We also confirm a logarithmic temperature dependence of the conductance, which is another typical signature of the Kondo correlation.
Although this signature does not appear in Fig.\,\ref{fig:device}(d), we find Kondo correlation also for the valley between the peak 1B and 2B, 3B and 4B, 7B and 8B by suitably tuning the coupling between the QD and the nearby reservoirs.
The phase behavior for Kondo correlated CPs has been investigated both theoretically \cite{Gerland2000, Hecht2009} and experimentally \cite{Takada2014, Takada2016}, and an extremely well agreement has been achieved.
In this experiment we investigate the phase behavior at temperatures well above the Kondo temperature ($T \gg T_{\rm K}$), which is expected to show the asymmetric behavior depending on the orbital parity relations with nearby CPs \cite{Hecht2009, Takada2014}.
Let us mention that the Coulomb diamond around the peaks 7A and 8A shows a slightly irregular feature.
This originates most probably from an impurity around the QD since such an impurit potential is less screened due to the low electron density of the QD.
On the other hand we do not observe significant influence from this effect around zero bias as can be seen in Fig.\,\ref{fig:device}c, where we perform phase measurements.

\begin{figure}[htbp]
	\includegraphics[width=0.48\textwidth]{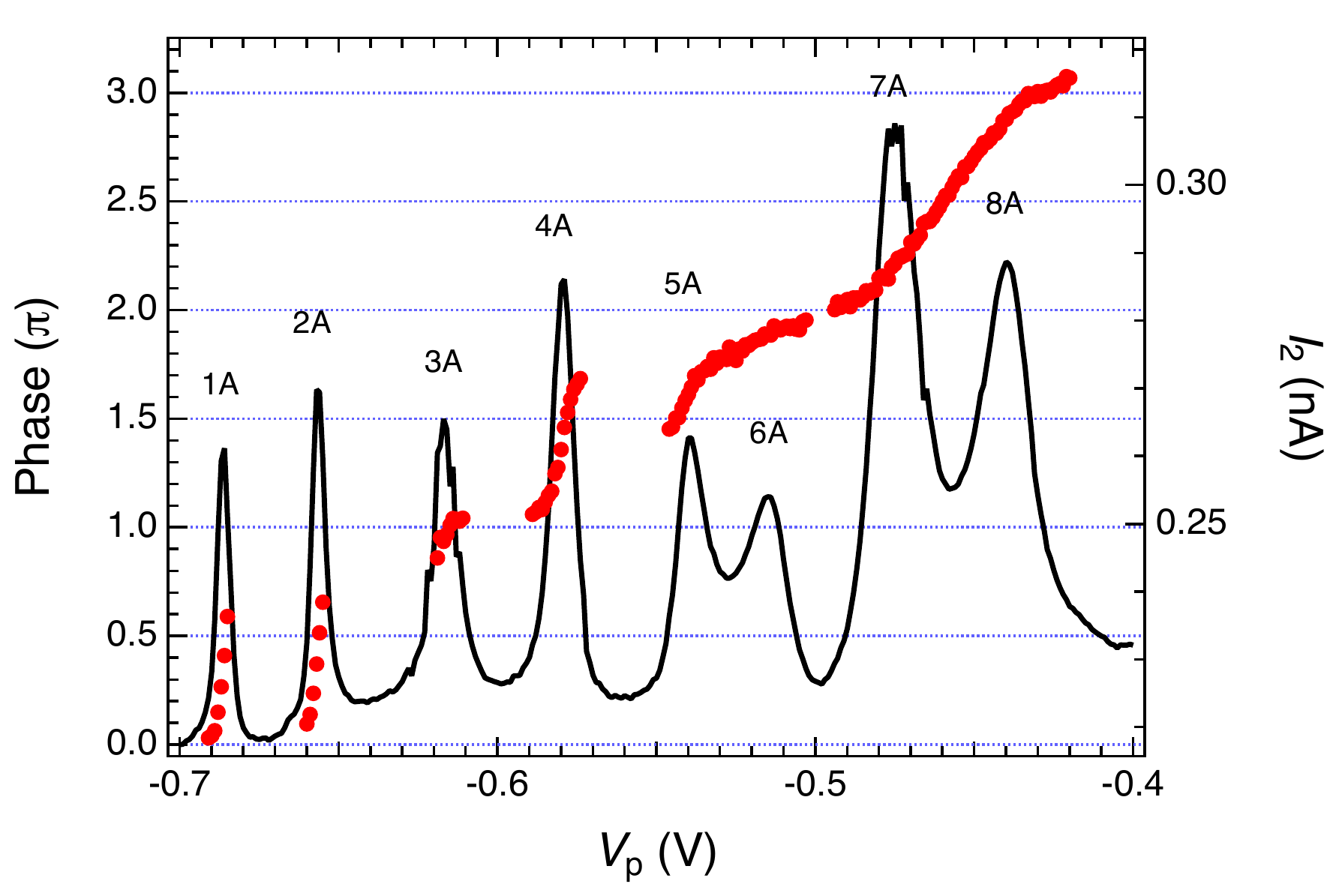}
	\caption{Transmission phase shift across the quantum dot in regime A. The phase shift (red circles) is plotted for the left axis and the current $I_{\rm 2}$ (black solid line) averaged over one oscillation period of the magnetic field is plotted for the right axis.\label{fig:PhaseShiftA}}
\end{figure}
For the phase measurements we retune the tunnel-coupled wires to be half beam splitters and measure the transmission phase shift across the QD.
This is done by recording the magnetic field dependence of both $I_{\rm 1}$ and $I_{\rm 2}$ at different plunger gate voltages $V_{\rm p}$ to change the number of electrons inside the QD \cite{Takada2014}.
We perform a complex fast Fourier transform of ($I_{\rm 1} - I_{\rm 2}$), which contains anti-phase components as a function of the magnetic field, to obtain the numerical value of the phase shift.
Let us first concentrate on regime A whose results are shown in Fig.\,\ref{fig:PhaseShiftA}.
Here we plot the phase shift and the current $I_{\rm 2}$ on the left and the right axis, respectively.
We only plot the phase data obtained from well defined anti-phase oscillations of $I_{\rm 1}$ and $I_{\rm 2}$ since the phase shift obtained from non-well defined anti-phase oscillations contains extra contributions from multiple path interferences \cite{TakadaAPL2015}.
In the Coulomb blockade region between CPs it is usually difficult to obtain clear anti-phase oscillations due to the small conductance.
This generally limits reliable data acquisition in close vicinity to the Coulomb blockade region and hence prevents the observation of the full phase shift of $\pi$ across a CP.
However we still obtain sufficiently large phase shifts to judge whether a phase lapse is present at each valley.

The phase evolution across the CPs shows a variety of different behaviors.
The monotonic phase evolution across two CPs of 5A and 6A,  7A and 8A is associated with the Kondo correlation \cite{Takada2016}.
The important feature is the absence of a phase lapse in the valleys between the peak 2A and 3A, 6A and 7A, which is a signature of the mesoscopic behavior \cite{Heiblum_phase_2005}.
This means that the mesoscopic behavior can be observed even if the QD is not in the few electron regime.
For this QD condition we estimate $\Gamma$ from Lorentzian fitting of the CPs and find that the crossover between single-level ($\Gamma < \delta$) and multi-level ($\Gamma > \delta$) transport regimes occurs between the peak 2A and 3A.
This result shows that near the crossover the phase behavior is still mesoscopic even for the multi-level transport regime.

We note that the observed phase shift contains a trivial phase shift from the modulation of the geometrical phase along the AB ring induced by $V_{\rm p}$.
We find it is sufficiently small ($\lesssim 5$ \% of $\pi$) compared to the total phase shift at each CP considering the capacitance of the gate.
For the phase measurements the tunnel coupling energy $\Gamma$ between the QD and the leads is tuned to be large enough to have large conductance through the QD over the swept range of $V_{\rm p}$ but small enough with respect to $U$ to avoid significant overlap between adjacent CPs.
The overlap between CPs prevents observation of a well defined Coulomb blockade between the CPs, and leads to large fluctuations of the electron number inside the QD and hence a much smaller phase shift compared to $\pi$ across each CP \cite{Yeyati1995}.
This makes judgement of the absence/presence of a phase lapse difficult.

\begin{figure}[htbp]
	\includegraphics[width=0.48\textwidth]{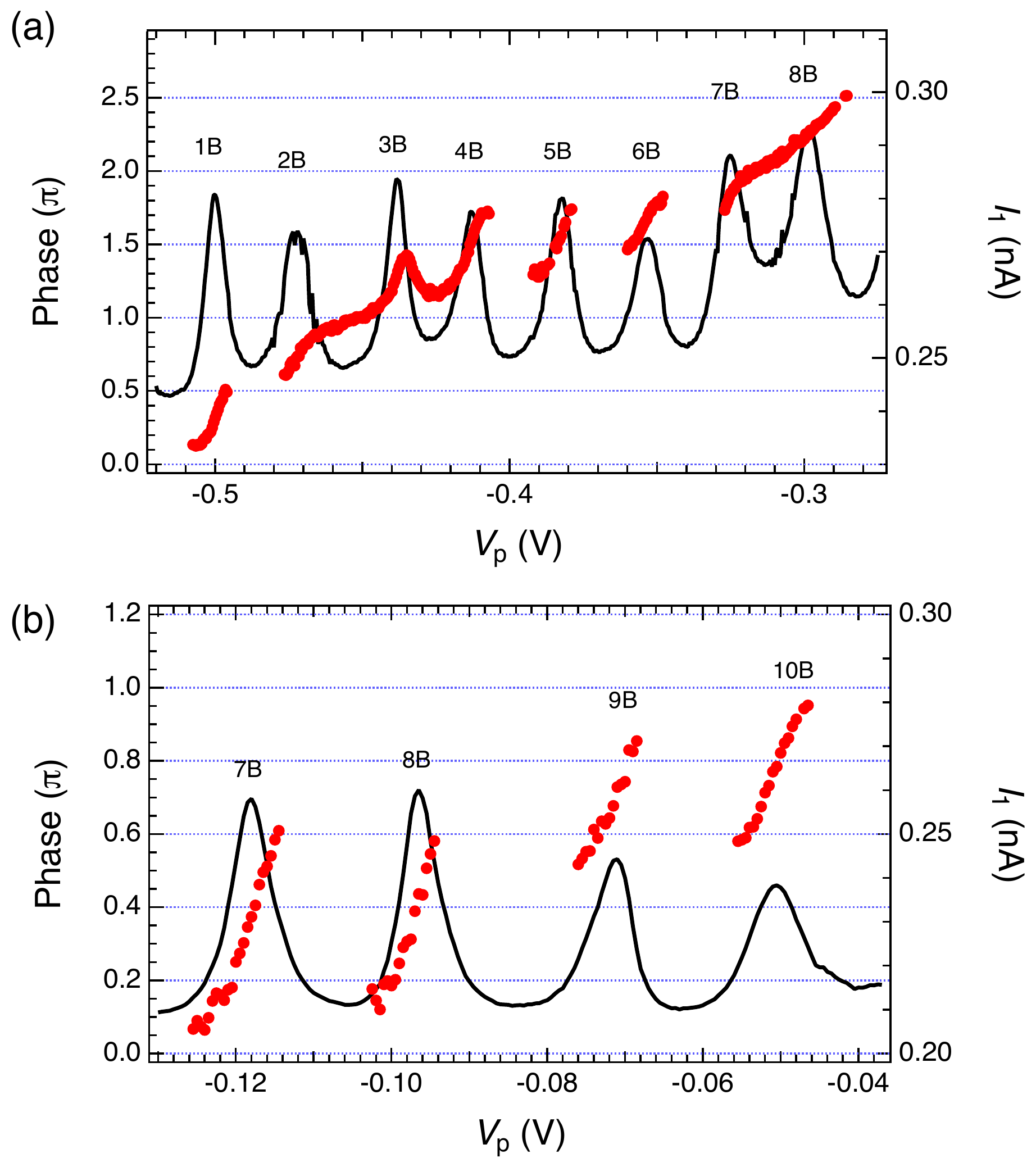}
	\caption{Transmission phase shift across the quantum dot in regime B. The phase shift (red circles) is plotted for the left axis and the current $I_{\rm 2}$ (black solid line) averaged over one oscillation period of the magnetic field is plotted for the right axis. In (b) the tunnel-coupling energy between the QD and the reservoir is tuned to be smaller than in (a). \label{fig:PhaseShiftB}}
\end{figure}
Next we measure the phase shift in regime B, which covers the larger QD.
In this regime the crossover between the two transport regimes occurs around the peak 1B.
Here we also observe a mesoscopic phase behavior as shown in Fig.\,\ref{fig:PhaseShiftB}a.
We confirm that there is no phase lapse in the valley between the peak 2B and 3B.
For the larger QD $\Gamma$ starts becoming large with respect to $U$ and the phase shift across each CP gets smaller compared to $\pi$.
In such a situation it becomes difficult to clearly judge whether there is a phase lapse or not in a valley.
On the other hand, the total phase shift from peak 5B to the peak 8B exceeds $\pi$ in Fig.\,\ref{fig:PhaseShiftB}a.
This result is inconsistent with the universal phase behavior.

We further increased the number of electrons in the QD to try to investigate the absence/presence of a phase lapse for the even larger QD.
However, since the total phase shift across each CP could not be made close to $\pi$ due to the limited tunability, this was not possible (see Fig.\,\ref{fig:PhaseShiftB}b).
We do not reach the universal phase behavior with the largest QD we could reach with this device.

\begin{figure}[htbp]
	\includegraphics[width=0.48\textwidth]{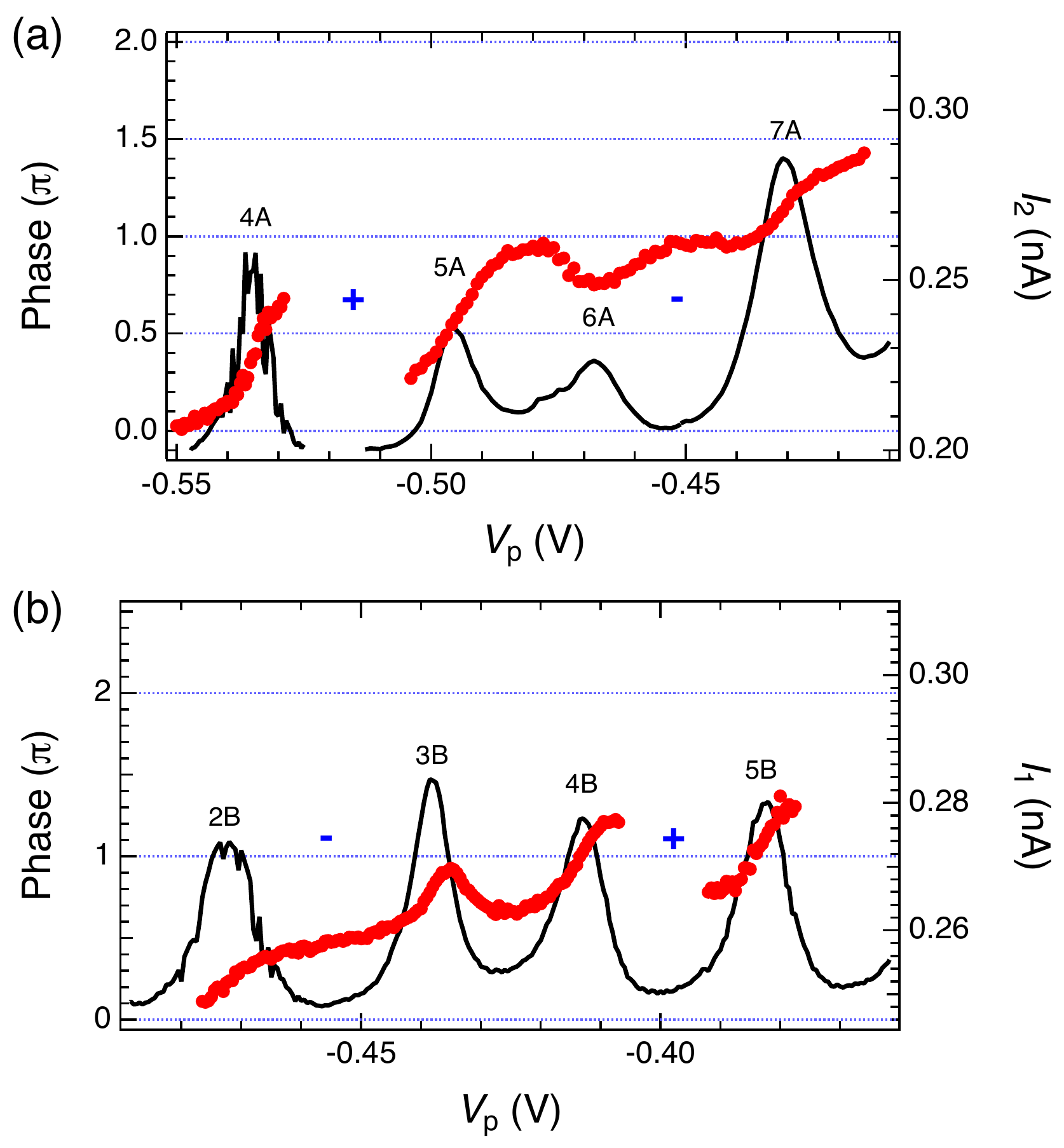}
	\caption{Transmission phase shift across Coulomb peaks with weak Kondo correlation ($T \gg T_{\rm K}$) with the phase shift across nearby Coulomb peaks for two different sets of the Coulomb peaks [(a) and (b)]. The parity relation between the orbital of the Kondo correlated Coulomb peaks and the orbital of the nearby Coulomb peaks, which is predicted from the appearance of the phase lapse, is indicated in the valley between the peaks. When the parity is same (opposite), it is indicated by $+$ ($-$). \label{fig:Parity}}
\end{figure}
Finally we show an indication that the absence/presence of the phase lapse is related with the orbital parity relation between nearby levels as theoretically expected.
This can be investigated from the asymmetric phase evolution across a pair of CPs showing weak Kondo correlation ($T \gg T_{\rm K}$) compared to the absence/presence of a phase lapse at the valley in the outer vicinity of the pair of CPs.
Here we focus on the phase behavior across the two pairs of CPs, 5A and 6A, 3B and 4B, which shows strong Kondo correlations under certain tuning conditions.
It has been theoretically shown that the phase evolution across two CPs with Kondo correlations can be asymmetric for $T \gg T_{\rm K}$, depending on the orbital parity relation with the other CPs nearby \cite{Hecht2009, Takada2014}.
When the symmetry of the orbital wave function responsible for the Kondo correlation and that of the adjacent CPs is the same on one side of the valley and is opposite on the other side, the phase evolution is larger across the peak closer to the valley with the same orbital parity and smaller across the other.
Such a phase evolution across the CPs with Kondo correlation (5A and 6A, 3B and 4B) at $T \gg T_{\rm K}$ is plotted in Fig\,\ref{fig:Parity} together with the phase evolution across the nearby CPs (4A and 7A, 2B and 5B).
In Fig\,\ref{fig:Parity}a the phase lapse appears between the CP 4A and 5A but not between the CP 6A and 7A.
According to Ref.\,\onlinecite{Lee1999}, the phase lapse appears in the valley when the orbital parity relation between nearby CPs is the same.
On the other hand, it does not appear when the parity is opposite.
Following this theory, the orbital parity relation should be the same for the valley between the CP 4A and 5A, and the opposite for the valley between the CP 6A and 7A.
Indeed, we find that the phase evolution is larger across the CP closer to the valley with the same orbital parity (5A) and smaller across the other (6A), consistent with theory.
A similar behavior is observed for regime B (Fig.\,\ref{fig:Parity}b) and hence corroborates this interpretation.
This result shows the connection between the appearance of a phase lapse and the asymmetry of the phase evolution at $T \gg T_{\rm K}$ as expected from Ref.\,\onlinecite{Hecht2009}.
Since calculations in Ref.\,\onlinecite{Hecht2009} include Friedel's sum rule relating an orbital parity with a phase lapse \cite{Lee1999}, this correspondence between the experimental results and the theoretical predictions in Ref.\,\onlinecite{Hecht2009} also supports the connection between the orbital parity relation and the appearance of a phase lapse.

In conclusion we have studied the transmission phase shift of an electron across a QD around the crossover between  single-level and multi-level transport.
We have found that the transmission phase shift around the crossover shows the ``predicted" mesoscopic behavior.
We also confirm that the experimentally observed relation between the appearance of phase lapses and the asymmetry of the phase evolution for CPs with weak Kondo correlation ($T \gg T_{\rm K}$) is consistent with theoretical calculations \cite{Hecht2009}.
This result supports the predicted connection between a phase lapse and an orbital parity relation \cite{Lee1999}.
Our results show that the universal phase lapse does not exist at least for a QD of a similar dimension, in contrast with the result reported in previous studies \cite{Heiblum_phase_2005}.
The question which remains is whether the universal phase laps appears for much larger QDs with well separated CPs while $\Gamma$ is clearly larger than $\delta$ \cite{Karrasch2007}.
Such a situation could not be investigated in our setup, neither in the setups of the previous works \cite{Schuster1997, Heiblum_phase_2005} as the QDs are too small.
This regime remains to be investigated.
Our observation shows the theoretically expected mesoscopic behavior of the phase lapse, which has been experimentally elusive for QDs in the non-few-electron regime.
This makes a significant step towards full understanding of a transmission phase behavior of a QD.

S. Takada acknowledges financial support from the European Unions Horizon 2020 research and innovation program under the Marie Sklodowska-Curie grant agreement No 654603.
M.Y. acknowledges financial support by Grant-in-Aid for Young Scientists A (No. 23684019), Grant-in-Aid for Challenging Exploratory Research (No. 25610070) and a Grant-in-Aid for Scientific Research A (No. 26247050).
C.B. acknowledges financial support from the French National Agency (ANR) in the frame of its program BLANC FLYELEC project No. anr-12BS10-001, as well as from DRECI-CNRS/JSPS (PRC 0677) international collaboration.
A.L. and A.D.W. acknowledge gratefully the support of Mercur Pr-2013-0001, DFG-TRR160, BMBF-Q.com-H 16KIS0109, and the DFH/UFA CDFA-05-06.
S. Tarucha acknowledges financial support by JSPS, Grant-in-Aid for Scientific Research S (No. 26220710), MEXT project for Developing Innovation Systems, and QPEC, the University of Tokyo.

\bibliography{/Users/Shintaro/Dropbox/Bibtex/library}

\begin{thebibliography}{29}%
\makeatletter
\providecommand \@ifxundefined [1]{%
 \@ifx{#1\undefined}
}%
\providecommand \@ifnum [1]{%
 \ifnum #1\expandafter \@firstoftwo
 \else \expandafter \@secondoftwo
 \fi
}%
\providecommand \@ifx [1]{%
 \ifx #1\expandafter \@firstoftwo
 \else \expandafter \@secondoftwo
 \fi
}%
\providecommand \natexlab [1]{#1}%
\providecommand \enquote  [1]{``#1''}%
\providecommand \bibnamefont  [1]{#1}%
\providecommand \bibfnamefont [1]{#1}%
\providecommand \citenamefont [1]{#1}%
\providecommand \href@noop [0]{\@secondoftwo}%
\providecommand \href [0]{\begingroup \@sanitize@url \@href}%
\providecommand \@href[1]{\@@startlink{#1}\@@href}%
\providecommand \@@href[1]{\endgroup#1\@@endlink}%
\providecommand \@sanitize@url [0]{\catcode `\\12\catcode `\$12\catcode
  `\&12\catcode `\#12\catcode `\^12\catcode `\_12\catcode `\%12\relax}%
\providecommand \@@startlink[1]{}%
\providecommand \@@endlink[0]{}%
\providecommand \url  [0]{\begingroup\@sanitize@url \@url }%
\providecommand \@url [1]{\endgroup\@href {#1}{\urlprefix }}%
\providecommand \urlprefix  [0]{URL }%
\providecommand \Eprint [0]{\href }%
\providecommand \doibase [0]{http://dx.doi.org/}%
\providecommand \selectlanguage [0]{\@gobble}%
\providecommand \bibinfo  [0]{\@secondoftwo}%
\providecommand \bibfield  [0]{\@secondoftwo}%
\providecommand \translation [1]{[#1]}%
\providecommand \BibitemOpen [0]{}%
\providecommand \bibitemStop [0]{}%
\providecommand \bibitemNoStop [0]{.\EOS\space}%
\providecommand \EOS [0]{\spacefactor3000\relax}%
\providecommand \BibitemShut  [1]{\csname bibitem#1\endcsname}%
\let\auto@bib@innerbib\@empty
\bibitem [{\citenamefont {Schuster}\ \emph {et~al.}(1997)\citenamefont
  {Schuster}, \citenamefont {Buks}, \citenamefont {Heiblum}, \citenamefont
  {Mahalu}, \citenamefont {Umansky},\ and\ \citenamefont
  {Shtrikman}}]{Schuster1997}%
  \BibitemOpen
  \bibfield  {author} {\bibinfo {author} {\bibfnamefont {R.}~\bibnamefont
  {Schuster}}, \bibinfo {author} {\bibfnamefont {E.}~\bibnamefont {Buks}},
  \bibinfo {author} {\bibfnamefont {M.}~\bibnamefont {Heiblum}}, \bibinfo
  {author} {\bibfnamefont {D.}~\bibnamefont {Mahalu}}, \bibinfo {author}
  {\bibfnamefont {V.}~\bibnamefont {Umansky}}, \ and\ \bibinfo {author}
  {\bibfnamefont {H.}~\bibnamefont {Shtrikman}},\ }\href {\doibase
  10.1038/385417a0} {\bibfield  {journal} {\bibinfo  {journal} {Nature}\
  }\textbf {\bibinfo {volume} {385}},\ \bibinfo {pages} {417} (\bibinfo {year}
  {1997})}\BibitemShut {NoStop}%
\bibitem [{\citenamefont {{Levy Yeyati}}\ \emph {et~al.}(1995)\citenamefont
  {{Levy Yeyati}}, \citenamefont {B{\"{u}}ttiker}, \citenamefont {Yeyati},\
  and\ \citenamefont {B{\"{u}}ttiker}}]{Yeyati1995}%
  \BibitemOpen
  \bibfield  {author} {\bibinfo {author} {\bibfnamefont {A.}~\bibnamefont
  {{Levy Yeyati}}}, \bibinfo {author} {\bibfnamefont {M.}~\bibnamefont
  {B{\"{u}}ttiker}}, \bibinfo {author} {\bibfnamefont {A.~L.}\ \bibnamefont
  {Yeyati}}, \ and\ \bibinfo {author} {\bibfnamefont {M.}~\bibnamefont
  {B{\"{u}}ttiker}},\ }\href {\doibase 10.1103/physrevb.52.r14360} {\bibfield
  {journal} {\bibinfo  {journal} {Phys. Rev. B}\ }\textbf {\bibinfo {volume}
  {52}},\ \bibinfo {pages} {R14360} (\bibinfo {year} {1995})}\BibitemShut
  {NoStop}%
\bibitem [{\citenamefont {Avinun-Kalish}\ \emph {et~al.}(2005)\citenamefont
  {Avinun-Kalish}, \citenamefont {Heiblum}, \citenamefont {Zarchin},
  \citenamefont {Mahalu},\ and\ \citenamefont {Umansky}}]{Heiblum_phase_2005}%
  \BibitemOpen
  \bibfield  {author} {\bibinfo {author} {\bibfnamefont {M.}~\bibnamefont
  {Avinun-Kalish}}, \bibinfo {author} {\bibfnamefont {M.}~\bibnamefont
  {Heiblum}}, \bibinfo {author} {\bibfnamefont {O.}~\bibnamefont {Zarchin}},
  \bibinfo {author} {\bibfnamefont {D.}~\bibnamefont {Mahalu}}, \ and\ \bibinfo
  {author} {\bibfnamefont {V.}~\bibnamefont {Umansky}},\ }\href@noop {}
  {\bibfield  {journal} {\bibinfo  {journal} {Nature (London)}\ }\textbf
  {\bibinfo {volume} {436}},\ \bibinfo {pages} {529} (\bibinfo {year}
  {2005})}\BibitemShut {NoStop}%
\bibitem [{\citenamefont {{-W. Lee}}(1999)}]{Lee1999}%
  \BibitemOpen
  \bibfield  {author} {\bibinfo {author} {\bibfnamefont {H.}~\bibnamefont {{-W.
  Lee}}},\ }\href {\doibase 10.1103/physrevlett.82.2358} {\bibfield  {journal}
  {\bibinfo  {journal} {Phys. Rev. Lett.}\ }\textbf {\bibinfo {volume} {82}},\
  \bibinfo {pages} {2358} (\bibinfo {year} {1999})}\BibitemShut {NoStop}%
\bibitem [{\citenamefont {Baltin}\ and\ \citenamefont
  {Gefen}(1999)}]{Baltin1999}%
  \BibitemOpen
  \bibfield  {author} {\bibinfo {author} {\bibfnamefont {R.}~\bibnamefont
  {Baltin}}\ and\ \bibinfo {author} {\bibfnamefont {Y.}~\bibnamefont {Gefen}},\
  }\href@noop {} {\bibfield  {journal} {\bibinfo  {journal} {Phys. Rev. Lett.}\
  }\textbf {\bibinfo {volume} {83}},\ \bibinfo {pages} {5094} (\bibinfo {year}
  {1999})}\BibitemShut {NoStop}%
\bibitem [{\citenamefont {Silvestrov}\ and\ \citenamefont
  {Imry}(2000)}]{Silvestrov2000}%
  \BibitemOpen
  \bibfield  {author} {\bibinfo {author} {\bibfnamefont {P.~G.}\ \bibnamefont
  {Silvestrov}}\ and\ \bibinfo {author} {\bibfnamefont {Y.}~\bibnamefont
  {Imry}},\ }\href {\doibase 10.1103/PhysRevLett.85.2565} {\bibfield  {journal}
  {\bibinfo  {journal} {Phys. Rev. Lett.}\ }\textbf {\bibinfo {volume} {85}},\
  \bibinfo {pages} {2565} (\bibinfo {year} {2000})}\BibitemShut {NoStop}%
\bibitem [{\citenamefont {{Levy Yeyati}}\ and\ \citenamefont
  {B{\"{u}}ttiker}(2000)}]{Yeyati2000}%
  \BibitemOpen
  \bibfield  {author} {\bibinfo {author} {\bibfnamefont {A.}~\bibnamefont
  {{Levy Yeyati}}}\ and\ \bibinfo {author} {\bibfnamefont {M.}~\bibnamefont
  {B{\"{u}}ttiker}},\ }\href {\doibase 10.1103/PhysRevB.62.7307} {\bibfield
  {journal} {\bibinfo  {journal} {Phys. Rev. B}\ }\textbf {\bibinfo {volume}
  {62}},\ \bibinfo {pages} {7307} (\bibinfo {year} {2000})}\BibitemShut
  {NoStop}%
\bibitem [{\citenamefont {Hackenbroich}(2001)}]{Hackenbroich2001}%
  \BibitemOpen
  \bibfield  {author} {\bibinfo {author} {\bibfnamefont {G.}~\bibnamefont
  {Hackenbroich}},\ }\href@noop {} {\bibfield  {journal} {\bibinfo  {journal}
  {Phys. Rep.}\ }\textbf {\bibinfo {volume} {343}},\ \bibinfo {pages} {463}
  (\bibinfo {year} {2001})}\BibitemShut {NoStop}%
\bibitem [{\citenamefont {Kim}\ \emph {et~al.}(2002)\citenamefont {Kim},
  \citenamefont {Cho}, \citenamefont {Kim},\ and\ \citenamefont
  {Ryu}}]{Kim2002a}%
  \BibitemOpen
  \bibfield  {author} {\bibinfo {author} {\bibfnamefont {T.-S.}\ \bibnamefont
  {Kim}}, \bibinfo {author} {\bibfnamefont {S.~Y.}\ \bibnamefont {Cho}},
  \bibinfo {author} {\bibfnamefont {C.~K.}\ \bibnamefont {Kim}}, \ and\
  \bibinfo {author} {\bibfnamefont {C.-M.}\ \bibnamefont {Ryu}},\ }\href
  {\doibase 10.1103/PhysRevB.65.245307} {\bibfield  {journal} {\bibinfo
  {journal} {Phys. Rev. B}\ }\textbf {\bibinfo {volume} {65}},\ \bibinfo
  {pages} {245307} (\bibinfo {year} {2002})}\BibitemShut {NoStop}%
\bibitem [{\citenamefont {Golosov}\ and\ \citenamefont
  {Gefen}(2006)}]{Golosov2006}%
  \BibitemOpen
  \bibfield  {author} {\bibinfo {author} {\bibfnamefont {D.~I.}\ \bibnamefont
  {Golosov}}\ and\ \bibinfo {author} {\bibfnamefont {Y.}~\bibnamefont
  {Gefen}},\ }\href {\doibase 10.1103/PhysRevB.74.205316} {\bibfield  {journal}
  {\bibinfo  {journal} {Phys. Rev. B}\ }\textbf {\bibinfo {volume} {74}},\
  \bibinfo {pages} {205316} (\bibinfo {year} {2006})}\BibitemShut {NoStop}%
\bibitem [{\citenamefont {Bertoni}\ and\ \citenamefont
  {Goldoni}(2007)}]{Bertoni2007a}%
  \BibitemOpen
  \bibfield  {author} {\bibinfo {author} {\bibfnamefont {A.}~\bibnamefont
  {Bertoni}}\ and\ \bibinfo {author} {\bibfnamefont {G.}~\bibnamefont
  {Goldoni}},\ }\href {\doibase 10.1103/PhysRevB.75.235318} {\bibfield
  {journal} {\bibinfo  {journal} {Phys. Rev. B}\ }\textbf {\bibinfo {volume}
  {75}},\ \bibinfo {pages} {235318} (\bibinfo {year} {2007})}\BibitemShut
  {NoStop}%
\bibitem [{\citenamefont {Karrasch}\ \emph {et~al.}(2007)\citenamefont
  {Karrasch}, \citenamefont {Hecht}, \citenamefont {Weichselbaum},
  \citenamefont {Oreg}, \citenamefont {von Delft},\ and\ \citenamefont
  {Meden}}]{Karrasch2007}%
  \BibitemOpen
  \bibfield  {author} {\bibinfo {author} {\bibfnamefont {C.}~\bibnamefont
  {Karrasch}}, \bibinfo {author} {\bibfnamefont {T.}~\bibnamefont {Hecht}},
  \bibinfo {author} {\bibfnamefont {A.}~\bibnamefont {Weichselbaum}}, \bibinfo
  {author} {\bibfnamefont {Y.}~\bibnamefont {Oreg}}, \bibinfo {author}
  {\bibfnamefont {J.}~\bibnamefont {von Delft}}, \ and\ \bibinfo {author}
  {\bibfnamefont {V.}~\bibnamefont {Meden}},\ }\href@noop {} {\bibfield
  {journal} {\bibinfo  {journal} {Phys. Rev. Lett.}\ }\textbf {\bibinfo
  {volume} {98}},\ \bibinfo {pages} {186802} (\bibinfo {year}
  {2007})}\BibitemShut {NoStop}%
\bibitem [{\citenamefont {Goldstein}\ \emph {et~al.}(2009)\citenamefont
  {Goldstein}, \citenamefont {Berkovits}, \citenamefont {Gefen},\ and\
  \citenamefont {Weidenm{\"{u}}ller}}]{Goldstein2009a}%
  \BibitemOpen
  \bibfield  {author} {\bibinfo {author} {\bibfnamefont {M.}~\bibnamefont
  {Goldstein}}, \bibinfo {author} {\bibfnamefont {R.}~\bibnamefont
  {Berkovits}}, \bibinfo {author} {\bibfnamefont {Y.}~\bibnamefont {Gefen}}, \
  and\ \bibinfo {author} {\bibfnamefont {H.~A.}\ \bibnamefont
  {Weidenm{\"{u}}ller}},\ }\href {\doibase 10.1103/PhysRevB.79.125307}
  {\bibfield  {journal} {\bibinfo  {journal} {Phys. Rev. B}\ }\textbf {\bibinfo
  {volume} {79}},\ \bibinfo {pages} {125307} (\bibinfo {year}
  {2009})}\BibitemShut {NoStop}%
\bibitem [{\citenamefont {Molina}\ \emph {et~al.}(2012)\citenamefont {Molina},
  \citenamefont {Jalabert}, \citenamefont {Weinmann},\ and\ \citenamefont
  {Jacquod}}]{Molina2012}%
  \BibitemOpen
  \bibfield  {author} {\bibinfo {author} {\bibfnamefont {R.~A.}\ \bibnamefont
  {Molina}}, \bibinfo {author} {\bibfnamefont {R.~A.}\ \bibnamefont
  {Jalabert}}, \bibinfo {author} {\bibfnamefont {D.}~\bibnamefont {Weinmann}},
  \ and\ \bibinfo {author} {\bibfnamefont {P.}~\bibnamefont {Jacquod}},\ }\href
  {\doibase 10.1103/PhysRevLett.108.076803} {\bibfield  {journal} {\bibinfo
  {journal} {Phys. Rev. Lett.}\ }\textbf {\bibinfo {volume} {108}},\ \bibinfo
  {pages} {076803} (\bibinfo {year} {2012})}\BibitemShut {NoStop}%
\bibitem [{\citenamefont {Sigrist}\ \emph {et~al.}(2004)\citenamefont
  {Sigrist}, \citenamefont {Fuhrer}, \citenamefont {Ihn}, \citenamefont
  {Ensslin}, \citenamefont {Ulloa}, \citenamefont {Wegscheider},\ and\
  \citenamefont {Bichler}}]{Sigrist2004}%
  \BibitemOpen
  \bibfield  {author} {\bibinfo {author} {\bibfnamefont {M.}~\bibnamefont
  {Sigrist}}, \bibinfo {author} {\bibfnamefont {A.}~\bibnamefont {Fuhrer}},
  \bibinfo {author} {\bibfnamefont {T.}~\bibnamefont {Ihn}}, \bibinfo {author}
  {\bibfnamefont {K.}~\bibnamefont {Ensslin}}, \bibinfo {author} {\bibfnamefont
  {S.~E.}\ \bibnamefont {Ulloa}}, \bibinfo {author} {\bibfnamefont
  {W.}~\bibnamefont {Wegscheider}}, \ and\ \bibinfo {author} {\bibfnamefont
  {M.}~\bibnamefont {Bichler}},\ }\href {\doibase
  10.1103/PhysRevLett.93.066802} {\bibfield  {journal} {\bibinfo  {journal}
  {Phys. Rev. Lett.}\ }\textbf {\bibinfo {volume} {93}},\ \bibinfo {pages}
  {066802} (\bibinfo {year} {2004})}\BibitemShut {NoStop}%
\bibitem [{\citenamefont {Aikawa}\ \emph {et~al.}(2004)\citenamefont {Aikawa},
  \citenamefont {Kobayashi}, \citenamefont {Sano}, \citenamefont {Katsumoto},\
  and\ \citenamefont {Iye}}]{Aikawa2004B}%
  \BibitemOpen
  \bibfield  {author} {\bibinfo {author} {\bibfnamefont {H.}~\bibnamefont
  {Aikawa}}, \bibinfo {author} {\bibfnamefont {K.}~\bibnamefont {Kobayashi}},
  \bibinfo {author} {\bibfnamefont {A.}~\bibnamefont {Sano}}, \bibinfo {author}
  {\bibfnamefont {S.}~\bibnamefont {Katsumoto}}, \ and\ \bibinfo {author}
  {\bibfnamefont {Y.}~\bibnamefont {Iye}},\ }\href {\doibase
  10.1143/JPSJ.73.3235} {\bibfield  {journal} {\bibinfo  {journal} {Journal of
  the Physical Society of Japan}\ }\textbf {\bibinfo {volume} {73}},\ \bibinfo
  {pages} {3235} (\bibinfo {year} {2004})}\BibitemShut {NoStop}%
\bibitem [{\citenamefont {Yacoby}\ \emph {et~al.}(1996)\citenamefont {Yacoby},
  \citenamefont {Schuster},\ and\ \citenamefont {Heiblum}}]{Yacoby1996}%
  \BibitemOpen
  \bibfield  {author} {\bibinfo {author} {\bibfnamefont {A.}~\bibnamefont
  {Yacoby}}, \bibinfo {author} {\bibfnamefont {R.}~\bibnamefont {Schuster}}, \
  and\ \bibinfo {author} {\bibfnamefont {M.}~\bibnamefont {Heiblum}},\ }\href
  {\doibase 10.1103/PhysRevB.53.9583} {\bibfield  {journal} {\bibinfo
  {journal} {Phys. Rev. B}\ }\textbf {\bibinfo {volume} {53}},\ \bibinfo
  {pages} {9583} (\bibinfo {year} {1996})}\BibitemShut {NoStop}%
\bibitem [{\citenamefont {Entin-Wohlman}\ \emph {et~al.}(2002)\citenamefont
  {Entin-Wohlman}, \citenamefont {Aharony}, \citenamefont {Imry}, \citenamefont
  {Levinson},\ and\ \citenamefont {Schiller}}]{Entin2002}%
  \BibitemOpen
  \bibfield  {author} {\bibinfo {author} {\bibfnamefont {O.}~\bibnamefont
  {Entin-Wohlman}}, \bibinfo {author} {\bibfnamefont {A.}~\bibnamefont
  {Aharony}}, \bibinfo {author} {\bibfnamefont {Y.}~\bibnamefont {Imry}},
  \bibinfo {author} {\bibfnamefont {Y.}~\bibnamefont {Levinson}}, \ and\
  \bibinfo {author} {\bibfnamefont {A.}~\bibnamefont {Schiller}},\ }\href
  {\doibase 10.1103/PhysRevLett.88.166801} {\bibfield  {journal} {\bibinfo
  {journal} {Phys. Rev. Lett.}\ }\textbf {\bibinfo {volume} {88}},\ \bibinfo
  {pages} {166801} (\bibinfo {year} {2002})}\BibitemShut {NoStop}%
\bibitem [{\citenamefont {Simon}\ \emph {et~al.}(2005)\citenamefont {Simon},
  \citenamefont {Entin-Wohlman},\ and\ \citenamefont {Aharony}}]{Simon2005}%
  \BibitemOpen
  \bibfield  {author} {\bibinfo {author} {\bibfnamefont {P.}~\bibnamefont
  {Simon}}, \bibinfo {author} {\bibfnamefont {O.}~\bibnamefont
  {Entin-Wohlman}}, \ and\ \bibinfo {author} {\bibfnamefont {A.}~\bibnamefont
  {Aharony}},\ }\href {\doibase 10.1103/PhysRevB.72.245313} {\bibfield
  {journal} {\bibinfo  {journal} {Phys. Rev. B}\ }\textbf {\bibinfo {volume}
  {72}},\ \bibinfo {pages} {245313} (\bibinfo {year} {2005})}\BibitemShut
  {NoStop}%
\bibitem [{\citenamefont {Takada}\ \emph {et~al.}(2015)\citenamefont {Takada},
  \citenamefont {Yamamoto}, \citenamefont {B{\"{a}}uerle}, \citenamefont
  {Watanabe}, \citenamefont {Ludwig}, \citenamefont {Wieck},\ and\
  \citenamefont {Tarucha}}]{TakadaAPL2015}%
  \BibitemOpen
  \bibfield  {author} {\bibinfo {author} {\bibfnamefont {S.}~\bibnamefont
  {Takada}}, \bibinfo {author} {\bibfnamefont {M.}~\bibnamefont {Yamamoto}},
  \bibinfo {author} {\bibfnamefont {C.}~\bibnamefont {B{\"{a}}uerle}}, \bibinfo
  {author} {\bibfnamefont {K.}~\bibnamefont {Watanabe}}, \bibinfo {author}
  {\bibfnamefont {A.}~\bibnamefont {Ludwig}}, \bibinfo {author} {\bibfnamefont
  {A.~D.}\ \bibnamefont {Wieck}}, \ and\ \bibinfo {author} {\bibfnamefont
  {S.}~\bibnamefont {Tarucha}},\ }\href {\doibase
  http://dx.doi.org/10.1063/1.4928035} {\bibfield  {journal} {\bibinfo
  {journal} {Appl. Phys. Lett.}\ }\textbf {\bibinfo {volume} {107}},\ \bibinfo
  {pages} {63101} (\bibinfo {year} {2015})}\BibitemShut {NoStop}%
\bibitem [{\citenamefont {Yamamoto}\ \emph {et~al.}(2012)\citenamefont
  {Yamamoto}, \citenamefont {Takada}, \citenamefont {B{\"{a}}uerle},
  \citenamefont {Watanabe}, \citenamefont {Wieck},\ and\ \citenamefont
  {Tarucha}}]{Yamamoto:2012fk}%
  \BibitemOpen
  \bibfield  {author} {\bibinfo {author} {\bibfnamefont {M.}~\bibnamefont
  {Yamamoto}}, \bibinfo {author} {\bibfnamefont {S.}~\bibnamefont {Takada}},
  \bibinfo {author} {\bibfnamefont {C.}~\bibnamefont {B{\"{a}}uerle}}, \bibinfo
  {author} {\bibfnamefont {K.}~\bibnamefont {Watanabe}}, \bibinfo {author}
  {\bibfnamefont {A.~D.}\ \bibnamefont {Wieck}}, \ and\ \bibinfo {author}
  {\bibfnamefont {S.}~\bibnamefont {Tarucha}},\ }\href {\doibase
  10.1038/nnano.2012.28} {\bibfield  {journal} {\bibinfo  {journal} {Nature
  Nanotech.}\ }\textbf {\bibinfo {volume} {7}},\ \bibinfo {pages} {247}
  (\bibinfo {year} {2012})}\BibitemShut {NoStop}%
\bibitem [{\citenamefont {Bautze}\ \emph {et~al.}(2014)\citenamefont {Bautze},
  \citenamefont {S{\"{u}}ssmeier}, \citenamefont {Takada}, \citenamefont
  {Groth}, \citenamefont {Meunier}, \citenamefont {Yamamoto}, \citenamefont
  {Tarucha}, \citenamefont {Waintal},\ and\ \citenamefont
  {B{\"{a}}uerle}}]{Tobias2014}%
  \BibitemOpen
  \bibfield  {author} {\bibinfo {author} {\bibfnamefont {T.}~\bibnamefont
  {Bautze}}, \bibinfo {author} {\bibfnamefont {C.}~\bibnamefont
  {S{\"{u}}ssmeier}}, \bibinfo {author} {\bibfnamefont {S.}~\bibnamefont
  {Takada}}, \bibinfo {author} {\bibfnamefont {C.}~\bibnamefont {Groth}},
  \bibinfo {author} {\bibfnamefont {T.}~\bibnamefont {Meunier}}, \bibinfo
  {author} {\bibfnamefont {M.}~\bibnamefont {Yamamoto}}, \bibinfo {author}
  {\bibfnamefont {S.}~\bibnamefont {Tarucha}}, \bibinfo {author} {\bibfnamefont
  {X.}~\bibnamefont {Waintal}}, \ and\ \bibinfo {author} {\bibfnamefont
  {C.}~\bibnamefont {B{\"{a}}uerle}},\ }\href {\doibase
  10.1103/PhysRevB.89.125432} {\bibfield  {journal} {\bibinfo  {journal} {Phys.
  Rev. B}\ }\textbf {\bibinfo {volume} {89}},\ \bibinfo {pages} {125432}
  (\bibinfo {year} {2014})},\ \Eprint {http://arxiv.org/abs/1312.5194}
  {arXiv:1312.5194} \BibitemShut {NoStop}%
\bibitem [{\citenamefont {Aharony}\ \emph {et~al.}(2014)\citenamefont
  {Aharony}, \citenamefont {Takada}, \citenamefont {Entin-Wohlman},
  \citenamefont {Yamamoto},\ and\ \citenamefont {Tarucha}}]{Aharony2014}%
  \BibitemOpen
  \bibfield  {author} {\bibinfo {author} {\bibfnamefont {A.}~\bibnamefont
  {Aharony}}, \bibinfo {author} {\bibfnamefont {S.}~\bibnamefont {Takada}},
  \bibinfo {author} {\bibfnamefont {O.}~\bibnamefont {Entin-Wohlman}}, \bibinfo
  {author} {\bibfnamefont {M.}~\bibnamefont {Yamamoto}}, \ and\ \bibinfo
  {author} {\bibfnamefont {S.}~\bibnamefont {Tarucha}},\ }\href {\doibase
  10.1088/1367-2630/16/8/083015} {\bibfield  {journal} {\bibinfo  {journal}
  {New Journal of Physics}\ }\textbf {\bibinfo {volume} {16}},\ \bibinfo
  {pages} {83015} (\bibinfo {year} {2014})}\BibitemShut {NoStop}%
\bibitem [{\citenamefont {Kondo}(1964)}]{Kondo1964}%
  \BibitemOpen
  \bibfield  {author} {\bibinfo {author} {\bibfnamefont {J.}~\bibnamefont
  {Kondo}},\ }\href {\doibase 10.1143/PTP.32.37} {\bibfield  {journal}
  {\bibinfo  {journal} {Progress of Theoretical Physics}\ }\textbf {\bibinfo
  {volume} {32}},\ \bibinfo {pages} {37} (\bibinfo {year} {1964})}\BibitemShut
  {NoStop}%
\bibitem [{\citenamefont {Goldhaber-Gordon}\ \emph {et~al.}(1998)\citenamefont
  {Goldhaber-Gordon}, \citenamefont {Shtrikman}, \citenamefont {Mahalu},
  \citenamefont {Abusch-Magder}, \citenamefont {Meirav},\ and\ \citenamefont
  {Kastner}}]{Gordon1998}%
  \BibitemOpen
  \bibfield  {author} {\bibinfo {author} {\bibfnamefont {D.}~\bibnamefont
  {Goldhaber-Gordon}}, \bibinfo {author} {\bibfnamefont {H.}~\bibnamefont
  {Shtrikman}}, \bibinfo {author} {\bibfnamefont {D.}~\bibnamefont {Mahalu}},
  \bibinfo {author} {\bibfnamefont {D.}~\bibnamefont {Abusch-Magder}}, \bibinfo
  {author} {\bibfnamefont {U.}~\bibnamefont {Meirav}}, \ and\ \bibinfo {author}
  {\bibfnamefont {M.~A.}\ \bibnamefont {Kastner}},\ }\href@noop {} {\bibfield
  {journal} {\bibinfo  {journal} {Nature}\ }\textbf {\bibinfo {volume} {391}},\
  \bibinfo {pages} {156} (\bibinfo {year} {1998})}\BibitemShut {NoStop}%
\bibitem [{\citenamefont {Gerland}\ \emph {et~al.}(2000)\citenamefont
  {Gerland}, \citenamefont {von Delft}, \citenamefont {Costi},\ and\
  \citenamefont {Oreg}}]{Gerland2000}%
  \BibitemOpen
  \bibfield  {author} {\bibinfo {author} {\bibfnamefont {U.}~\bibnamefont
  {Gerland}}, \bibinfo {author} {\bibfnamefont {J.}~\bibnamefont {von Delft}},
  \bibinfo {author} {\bibfnamefont {T.~A.}\ \bibnamefont {Costi}}, \ and\
  \bibinfo {author} {\bibfnamefont {Y.}~\bibnamefont {Oreg}},\ }\href {\doibase
  10.1103/PhysRevLett.84.3710} {\bibfield  {journal} {\bibinfo  {journal}
  {Phys. Rev. Lett.}\ }\textbf {\bibinfo {volume} {84}},\ \bibinfo {pages}
  {3710} (\bibinfo {year} {2000})}\BibitemShut {NoStop}%
\bibitem [{\citenamefont {Hecht}\ \emph {et~al.}(2009)\citenamefont {Hecht},
  \citenamefont {Weichselbaum}, \citenamefont {Oreg},\ and\ \citenamefont {von
  Delft}}]{Hecht2009}%
  \BibitemOpen
  \bibfield  {author} {\bibinfo {author} {\bibfnamefont {T.}~\bibnamefont
  {Hecht}}, \bibinfo {author} {\bibfnamefont {A.}~\bibnamefont {Weichselbaum}},
  \bibinfo {author} {\bibfnamefont {Y.}~\bibnamefont {Oreg}}, \ and\ \bibinfo
  {author} {\bibfnamefont {J.}~\bibnamefont {von Delft}},\ }\href {\doibase
  10.1103/PhysRevB.80.115330} {\bibfield  {journal} {\bibinfo  {journal} {Phys.
  Rev. B}\ }\textbf {\bibinfo {volume} {80}},\ \bibinfo {pages} {115330}
  (\bibinfo {year} {2009})}\BibitemShut {NoStop}%
\bibitem [{\citenamefont {Takada}\ \emph {et~al.}(2014)\citenamefont {Takada},
  \citenamefont {B{\"{a}}uerle}, \citenamefont {Yamamoto}, \citenamefont
  {Watanabe}, \citenamefont {Hermelin}, \citenamefont {Meunier}, \citenamefont
  {Alex}, \citenamefont {Weichselbaum}, \citenamefont {von Delft},
  \citenamefont {Ludwig}, \citenamefont {Wieck},\ and\ \citenamefont
  {Tarucha}}]{Takada2014}%
  \BibitemOpen
  \bibfield  {author} {\bibinfo {author} {\bibfnamefont {S.}~\bibnamefont
  {Takada}}, \bibinfo {author} {\bibfnamefont {C.}~\bibnamefont
  {B{\"{a}}uerle}}, \bibinfo {author} {\bibfnamefont {M.}~\bibnamefont
  {Yamamoto}}, \bibinfo {author} {\bibfnamefont {K.}~\bibnamefont {Watanabe}},
  \bibinfo {author} {\bibfnamefont {S.}~\bibnamefont {Hermelin}}, \bibinfo
  {author} {\bibfnamefont {T.}~\bibnamefont {Meunier}}, \bibinfo {author}
  {\bibfnamefont {A.}~\bibnamefont {Alex}}, \bibinfo {author} {\bibfnamefont
  {A.}~\bibnamefont {Weichselbaum}}, \bibinfo {author} {\bibfnamefont
  {J.}~\bibnamefont {von Delft}}, \bibinfo {author} {\bibfnamefont
  {A.}~\bibnamefont {Ludwig}}, \bibinfo {author} {\bibfnamefont {A.~D.}\
  \bibnamefont {Wieck}}, \ and\ \bibinfo {author} {\bibfnamefont
  {S.}~\bibnamefont {Tarucha}},\ }\href {\doibase
  10.1103/PhysRevLett.113.126601} {\bibfield  {journal} {\bibinfo  {journal}
  {Phys. Rev. Lett.}\ }\textbf {\bibinfo {volume} {113}},\ \bibinfo {pages}
  {126601} (\bibinfo {year} {2014})},\ \Eprint {http://arxiv.org/abs/1311.6884}
  {arXiv:1311.6884} \BibitemShut {NoStop}%
\bibitem [{\citenamefont {Takada}\ \emph {et~al.}(2016)\citenamefont {Takada},
  \citenamefont {Yamamoto}, \citenamefont {B{\"{a}}uerle}, \citenamefont
  {Alex}, \citenamefont {von Delft}, \citenamefont {Ludwig}, \citenamefont
  {Wieck},\ and\ \citenamefont {Tarucha}}]{Takada2016}%
  \BibitemOpen
  \bibfield  {author} {\bibinfo {author} {\bibfnamefont {S.}~\bibnamefont
  {Takada}}, \bibinfo {author} {\bibfnamefont {M.}~\bibnamefont {Yamamoto}},
  \bibinfo {author} {\bibfnamefont {C.}~\bibnamefont {B{\"{a}}uerle}}, \bibinfo
  {author} {\bibfnamefont {A.}~\bibnamefont {Alex}}, \bibinfo {author}
  {\bibfnamefont {J.}~\bibnamefont {von Delft}}, \bibinfo {author}
  {\bibfnamefont {A.}~\bibnamefont {Ludwig}}, \bibinfo {author} {\bibfnamefont
  {A.~D.}\ \bibnamefont {Wieck}}, \ and\ \bibinfo {author} {\bibfnamefont
  {S.}~\bibnamefont {Tarucha}},\ }\href {\doibase 10.1103/PhysRevB.94.081303}
  {\bibfield  {journal} {\bibinfo  {journal} {Phys. Rev. B}\ }\textbf {\bibinfo
  {volume} {94}},\ \bibinfo {pages} {081303(R)} (\bibinfo {year}
  {2016})}\BibitemShut {NoStop}%
\end{thebibliography}%

\end{document}